\documentclass[aps,prl,amsmath,amssymb,footinbib,showpacs,twocolumn,superscriptaddress]{revtex4-2}
\usepackage{amsmath}
\usepackage{amssymb}
\usepackage{amsthm}
\usepackage{setspace}
\usepackage{graphicx}
\usepackage{braket}
\usepackage{mathrsfs}
\usepackage{float}
\usepackage[colorlinks = true,linkcolor = blue,urlcolor  = blue,citecolor = blue,anchorcolor = blue]{hyperref}
\usepackage[utf8]{inputenc}
\usepackage[english]{babel}
\usepackage{bm}

\begin{document}

\title{Revisiting magnetotransport in Weyl semimetals}

\author{G. Sharma}
\affiliation{School of Basic Sciences, Indian Institute of Technology Mandi, Mandi 175005, India}
\author{Snehashish Nandy}
\affiliation{Department of Physics, University of Virginia, Charlottesville, Virginia 22904, USA}
\author{Karthik V. Raman}
\affiliation{Tata Institute of Fundamental Research, Hyderabad, Telangana 500046, India}
\author{Sumanta Tewari}
\affiliation{Department of Physics and Astronomy, Clemson University, Clemson, South Carolina 29634, USA}

\begin{abstract}
%The importance of internode scattering contributing to positive longitudinal magnetoconductance (LMC) in Weyl semimetals (WSMs) via the chiral anomaly (CA), pointed out by Son and Spivak~\cite{son2013chiral}, was followed by 
A series of recent papers have claimed that intranode scattering, \textit{alone}, can contribute to positive longitudinal magnetoconductance (LMC) due to chiral anomaly (CA) in Weyl semimetals (WSMs).
We revisit the problem of CA induced LMC in WSMs, and show that intranode scattering, by itself, does not result in enhancement of LMC. In the limit of zero internode scattering, chiral charge must remain conserved, which is shown to actually decrease LMC. Only in the presence of a non-zero internode scattering (however weak), one obtains a positive LMC due to non-conservation of chiral charge. Even weak internode scattering suffices in generating positive LMC, since it redistributes charges across both the nodes, although on a time scale  larger than that of the intranode scattering.
%Our work, therefore, reconciles the Landau level and the semiclassical Boltzmann picture, showing that CA induced positive LMC is indeed only an internode phenomena. 
Furthermore, our calculations reveal that, in contrast to recent works, in inhomogeneous WSMs strain induced axial magnetic field $B_5$, by itself, leads to negative longitudinal magnetoconductance and a negative planar Hall conductance. %of opposite sign as compared to an external real magnetic field.
\end{abstract}

\maketitle

\textit{Introduction:}
Chiral anomaly (CA) finds its genesis in high-energy physics~\cite{adler1969axial,bell1969pcac}, whereby the left/right-handed Weyl fermions are not conserved in the presence of non-orthogonal electric and magnetic fields. The anomaly has resurged in Weyl semimetals (WSMs), and has been of great interest in the condensed-matter community over the last decade~\cite{armitage2018weyl,volovik2003universe,nielsen1981no,nielsen1983adler,wan2011topological,xu2011chern,zyuzin2012weyl,son2013chiral,goswami2013axionic,goswami2015axial,zhong2015optical,kim2014boltzmann,lundgren2014thermoelectric,cortijo2016linear,sharma2016nernst,zyuzin2017magnetotransport,das2019berry,kundu2020magnetotransport,knoll2020negative,sharma2020sign}. Since Weyl semimetals host Weyl fermions as quasiparticle excitations, chiral anomaly is expected to occur in these materials in the presence of external electromagnetic fields. Positive longitudinal magnetoconductance (LMC)~\cite{son2013chiral}, and the planar Hall effect (PHE)~\cite{nandy2017chiral} are some key signatures to identify the manifestation of chiral anomaly in Weyl semimetals. Significant efforts have recently been devoted to capture the behavior of these anomaly induced conductivities in WSMs~\cite{zyuzin2012weyl,son2013chiral,goswami2013axionic,goswami2015axial,zhong2015optical,kim2014boltzmann,lundgren2014thermoelectric,cortijo2016linear,sharma2016nernst,zyuzin2017magnetotransport,das2019berry,kundu2020magnetotransport,knoll2020negative,sharma2020sign}.

The calculation of conductivity via the linear response formalism~\cite{mahan20089} inherently assumes a timescale $\tau_{\phi}$, that can be interpreted as arising from interactions between the system and the external electric field that inelastically exchange energy at a rate $1/\tau_\phi$. The rate is assumed to be ideally zero, or in other words, $\tau_\phi$ is assumed to be the largest of all relevant timescales.
In the context of weakly disordered Weyl semimetals, when Landau quantization is relevant at high magnetic fields, chiral anomaly manifests itself by a positive contribution to the longitudinal magnetoconductance, i.e., $\mathbf{j}\propto B(\mathbf{E} \cdot \mathbf{B})$, where $\mathbf{E}$ and $\mathbf{B}$ are the applied electric and magnetic fields. The current in this case is limited by the internode scattering time ($\tau_\mathrm{inter}$), which corresponds to a timescale at which quasiparticles scatter across the nodes and switch their chirality. 
Therefore, chiral charge is not conserved, but the global charge remains conserved. For linear response formalism to work, $\tau_\mathrm{inter}$ must be much less than $\tau_\phi$.
Unfortunately, this approach breaks down if one considers intranode scattering as the \textit{only} dominant scattering mechanism, since in this case the chiral charge must remain conserved along with the global charge. 

In the limit of weak magnetic field, Son and Spivak~\cite{son2013chiral} highlighted the importance of internode scattering induced positive LMC in WSMs via the semiclassical Boltzmann approach. Despite this, a series of recent papers
suggested that intranode scattering, \textit{alone},  can give rise to positive longitudinal magnetoconductivity via a $\mathbf{E}\cdot\mathbf{B}$ force term incorporated in the semiclassical equations of motion~\cite{kim2014boltzmann,lundgren2014thermoelectric,cortijo2016linear,sharma2016nernst,zyuzin2017magnetotransport,das2019berry,kundu2020magnetotransport,knoll2020negative}. It is therefore implied that positive LMC manifests in WSMs even in the limit when  $\tau_\mathrm{inter}/\tau_\mathrm{intra}\rightarrow\infty$ (where $\tau_\mathrm{intra}$ is the intravalley scattering time). Importantly, these works also make no distinction between the parameter regimes $\tau_\mathrm{intra}\ll\tau_\phi\ll\tau_\mathrm{inter}$, and $\tau_\mathrm{intra}\ll\tau_\mathrm{inter}\ll\tau_\phi$. 
The distinction between the two cases actually has drastic consequences, as chiral charge is conserved in the former, but the latter indicates global charge conservation, which occurs on a timescale larger than the intravalley timescale $\tau_\mathrm{intra}$, but smaller than $\tau_\phi$. The claim that intranode scattering, alone, can positively increase LMC, inherently assumes that  $\tau_\mathrm{intra}\ll\tau_\phi\ll\tau_\mathrm{inter}$.

In this Letter, we show that the intranode scattering {alone} cannot yield positive LMC in WSMs, as this result is inconsistent with chiral charge conservation. We begin by first showing that semiclassical calculations of LMC with a momentum-independent scattering time, as is assumed in the recent works~\cite{kim2014boltzmann,lundgren2014thermoelectric,cortijo2016linear,sharma2016nernst,zyuzin2017magnetotransport,das2019berry,kundu2020magnetotransport}, violate chiral charge conservation. Chiral charge conservation is shown to be consistent only with a momentum-dependent scattering rate, which, very importantly, yields a $negative$ LMC. We therefore show that \textit{internode scattering is necessary} to obtain positive LMC in Weyl semimetals, as opposed to the recent works. Even very weak internode scattering drives the system to show positive LMC, and only sufficiently strong intervalley scattering beyond a critical strength switches the sign to a negative LMC~\cite{knoll2020negative}. Chiral anomaly is therefore an internode phenomena in WSMs, even in the weak-$B$ limit where the semiclassical formalism is valid, consistent with the Landau-level picture. 
Our results also have important consequences in the context of inhomogeneous WSMs (IWSMs), where we show that strain, alone, leads to a negative longitudinal magnetoconductance, and contributes a planar Hall conductance which is of opposite sign to that of an external magnetic field. These results, which can be experimentally verified, are also in contrast to some other recent papers~\cite{grushin2016inhomogeneous,ghosh2020chirality}. 

\textit{Inconsistencies of momentum-independent scattering time:}
For a single isolated Weyl node $H_\mathbf{k}=\chi \hbar v_F \mathbf{k}\cdot\boldsymbol{\sigma}$, the steady-state Boltzmann equation takes the following form~\cite{SI} 
\begin{align}
    e \mathcal{D^\chi}_\mathbf{k} \left(-\frac{\partial f_0}{\partial \epsilon_\mathbf{k}}\right) \left(\mathbf{v}^\chi_\mathbf{k}+ \frac{e}{\hbar}\mathbf{B} (\Omega^\chi_\mathbf{k}\cdot\mathbf{v}^\chi_\mathbf{k})\right)\cdot\mathbf{E} = \mathcal{I}_\mathrm{coll}\{f_\mathbf{k}\}
    \label{Eq:Boltz1}
\end{align}
where $\mathbf{v}^\chi_\mathbf{k}$ is the semiclassical band-velocity, $\mathbf{E}$ and $\mathbf{B}$ are external electric and magnetic fields, respectively, $\Omega_\mathbf{k}^\chi$ is the Berry curvature, $f_0$ is the equalibrium Boltzmann distribution, $\epsilon_\mathbf{k}$ is the energy, and $\mathcal{D}^\chi_\mathbf{k}=(1+e \mathbf{B} \cdot \Omega^\chi_\mathbf{k}/\hbar)^{-1}$.
Within the relaxation-time approximation, the collision integral $\mathcal{I}_\mathrm{coll}\{f_\mathbf{k}\}$ is simply 
\begin{align}
    \mathcal{I}_\mathrm{coll}\{f_\mathbf{k}\} = -\frac{\delta f^\chi_\mathbf{k}}{\tau_\mathbf{k}},
\end{align}
where $\delta f^\chi_\mathbf{k}$ is the deviation from the equilibrium. Particle conservation implies that 
\begin{align}
\sum\limits_\mathbf{k} \delta f^\chi_\mathbf{k} = 0
\label{Eq:particle_consv}
\end{align}

In the geometry when $\mathbf{E}\parallel\mathbf{B}\parallel\hat{z}$, the equation for particle conservation, Eq.~\ref{Eq:particle_consv}, reduces to the following form~\cite{SI}: 
\begin{align}
    \int {\tau^\chi(\theta) \left({v}^\chi_{z}+ \frac{e}{\hbar}{B} (\Omega^\chi_\mathbf{k}\cdot\mathbf{v}^\chi_\mathbf{k}) \right) \frac{k^3(\theta) \sin\theta}{|\mathbf{v}^\chi_\mathbf{k}\cdot\mathbf{k}|} d\theta} = 0. 
    \label{Eq:particle_consv_2}
\end{align}
Here all the quantities in the integrand are evaluated on the Fermi surface at zero temperature. A simplifying approach often employed is to assume that the scattering time is independent of momentum $\mathbf{k}$, i.e., $\tau^\chi(\theta) = \tau^\chi$~\cite{kim2014boltzmann,lundgren2014thermoelectric,cortijo2016linear,sharma2016nernst,zyuzin2017magnetotransport,das2019berry,kundu2020magnetotransport}. However, it can be  easily verified from Eq.~\ref{Eq:particle_consv_2} that when $\tau^\chi(\theta)$ is independent of the polar angle $\theta$, the L.H.S. of the equation does not reduce to zero! Even if we ignore the contribution due to the orbital magnetic moment, the integral above still turns out to be non-zero. 

For the case with a pair of Weyl nodes of opposite chiralities, if the internode scattering is assumed to be strictly zero, then particle number must be conserved on individual Weyl nodes, and thus, the above result of non-applicability of momentum independent scattering time is valid even for the case of two Weyl nodes. %as long as internode scattering is completely removed from the picture. 
A momentum-independent scattering time is thus inconsistent with particle number conservation, at least in the limit of infinite internode scattering time. 

\begin{figure}
    \centering
    \includegraphics[width=\columnwidth]{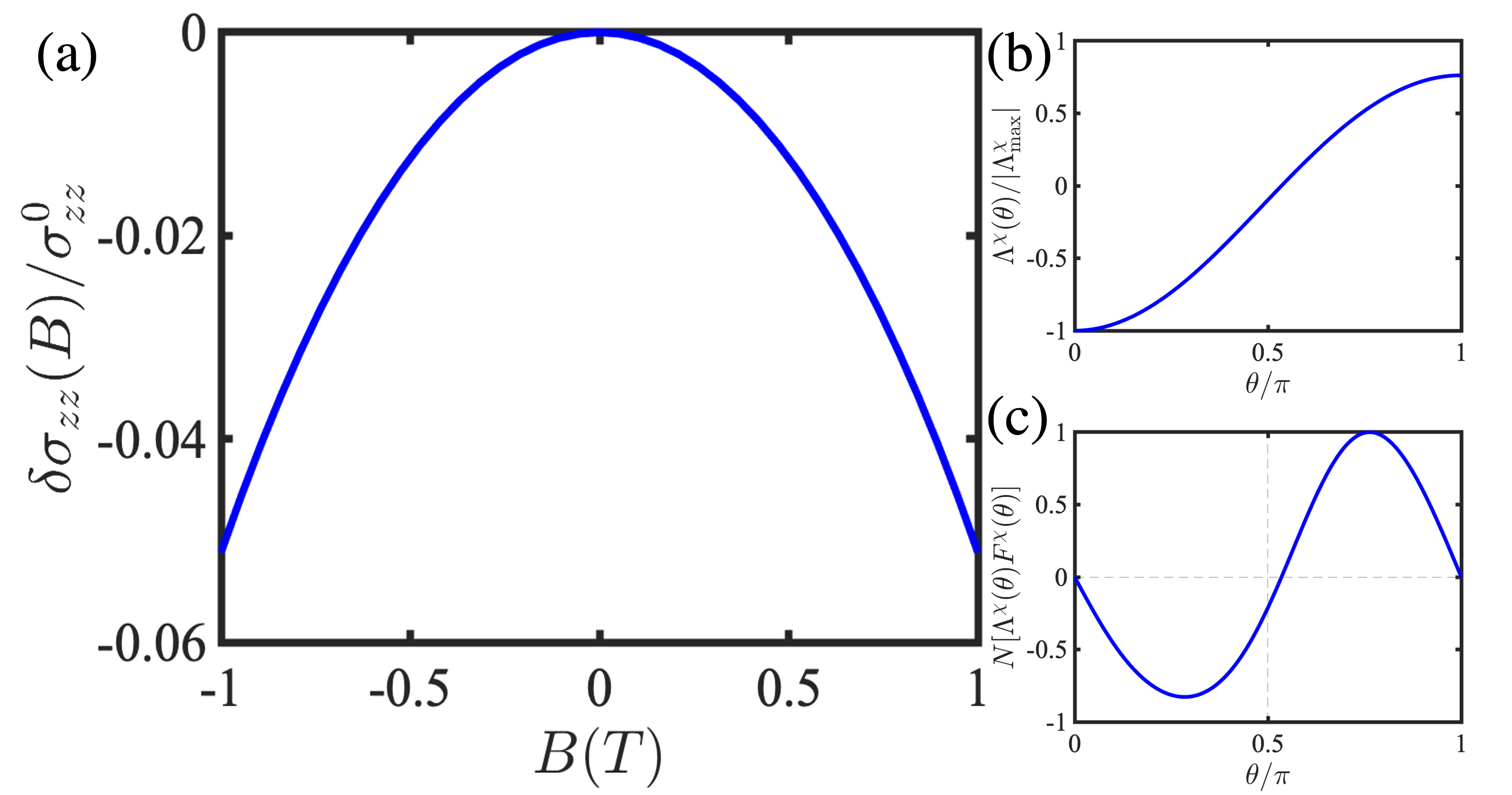}
    \caption{(a) LMC for an isolated Weyl node is always negative due to chiral charge conservation. (b)  $\Lambda^\chi(\theta)$ between $0$ and $\pi$ at $B=1T$. (c) The corresponding normalized deviation in the distribution function (proportional to $\delta f^\chi_\mathbf{k}$), which integrates to zero. Here we choose $\chi=1$.}
    \label{fig:one_node_1}
\end{figure}
\textit{LMC for zero internode scattering:} When internode scattering is zero, it suffices to calculate the result for a single isolated Weyl node, and the contribution from both the nodes can be simply summed over. Since we have shown that the momentum-independent relaxation time is not valid, we instead choose the collision integral in Eq.~\ref{Eq:Boltz1} to be:
\begin{align}
    \mathcal{I}_\mathrm{coll}\{f_\mathbf{k}\}=\sum\limits_{\mathbf{k}'} \left(\Lambda^\chi_\mathbf{k} - \Lambda^\chi_{\mathbf{k}'}\right) W_{\mathbf{k}\mathbf{k}'}.
\end{align}
Here the scattering rate $W_{\mathbf{k}\mathbf{k}'}$ must not be taken to be independent of momenta, and without loss of generality, the unknown function $\Lambda^\chi_\mathbf{k}$ can be assumed to be 
\begin{align}
    \Lambda^\chi_\mathbf{k} = (f^\chi_\mathbf{k} - h_\mathbf{k}^\chi) \tau_\mathbf{k}
\end{align}
where $h^\chi_\mathbf{k} = \mathcal{D^\chi}_\mathbf{k} \left({v}^\chi_z+ {e}{\hbar^{-1}}{B} (\Omega^\chi_\mathbf{k}\cdot\mathbf{v}^\chi_\mathbf{k})\right)$, $\tau^{-1}_\mathbf{k} = \sum_{\mathbf{k}'} W_{\mathbf{k}\mathbf{k}'}$, and the function $f_\mathbf{k}^\chi$ is the unknown. In the first Born approximation, and in the simplest case of point-like disorder, the scattering rate is evaluated to be
\begin{align}
    W_{\mathbf{k}\mathbf{k}'} &= \frac{2\pi}{\hbar} |U|^2 \delta(\epsilon_\mathbf{k} - \epsilon_F) \times \nonumber\\&(1+\cos\theta\cos\theta' + \sin\theta\sin\theta' \cos(\phi-\phi')),
    \label{Eq_W_1}
\end{align}
where $U$ is the strength of the disorder in appropriate units. We emphasize that the explicit dependence on the direction of momenta, i.e., the polar and azimuthal angles, which comes from the chirality of the Weyl fermion wavefunction, survives even in the absence of momentum-dependent disorder strength $U$. The Boltzmann equation reduces to 
\begin{align}
    f^\chi(\theta) =  \int{F^\chi(\theta')\tau(\theta) (1+\cos\theta\cos\theta') (f^\chi(\theta') - h^\chi(\theta')) d\theta'},
\end{align}
where $F^\chi(\theta) = k^3(\theta) (\mathcal{D}^{\chi}_\theta)^{-1} \sin\theta  |\mathbf{v}_\mathbf{k}\cdot\mathbf{k}|^{-1}$. Again, all the quantities in the integrand above are evaluated on the Fermi surface. To this end, incorporating particle conservation, along with the following ansatz solves the above Boltzmann transport equation:
\begin{align}
    \Lambda^\chi(\theta) = \tau(\theta) (a^\chi + b^\chi \cos\theta - h^\chi(\theta)),
\end{align}
 Finally, the current due to the deviation in equilibrium is evaluated as: 
\begin{align}
    \mathbf{j}^\chi = -e\sum\limits_{\mathbf{k}}{\dot{\mathbf{r}}^\chi (\delta f^\chi_\mathbf{k})},
\end{align}
and the longitudinal conductivity $\sigma_{zz}(B)$ can be evaluated. We define $\delta\sigma_{zz}(B) = \sigma_{zz}(B) - \sigma^0_{zz}$, where $\sigma^0_{zz}$ is the zero field conductivity,

\begin{figure}
    \centering
    \includegraphics[width=\columnwidth]{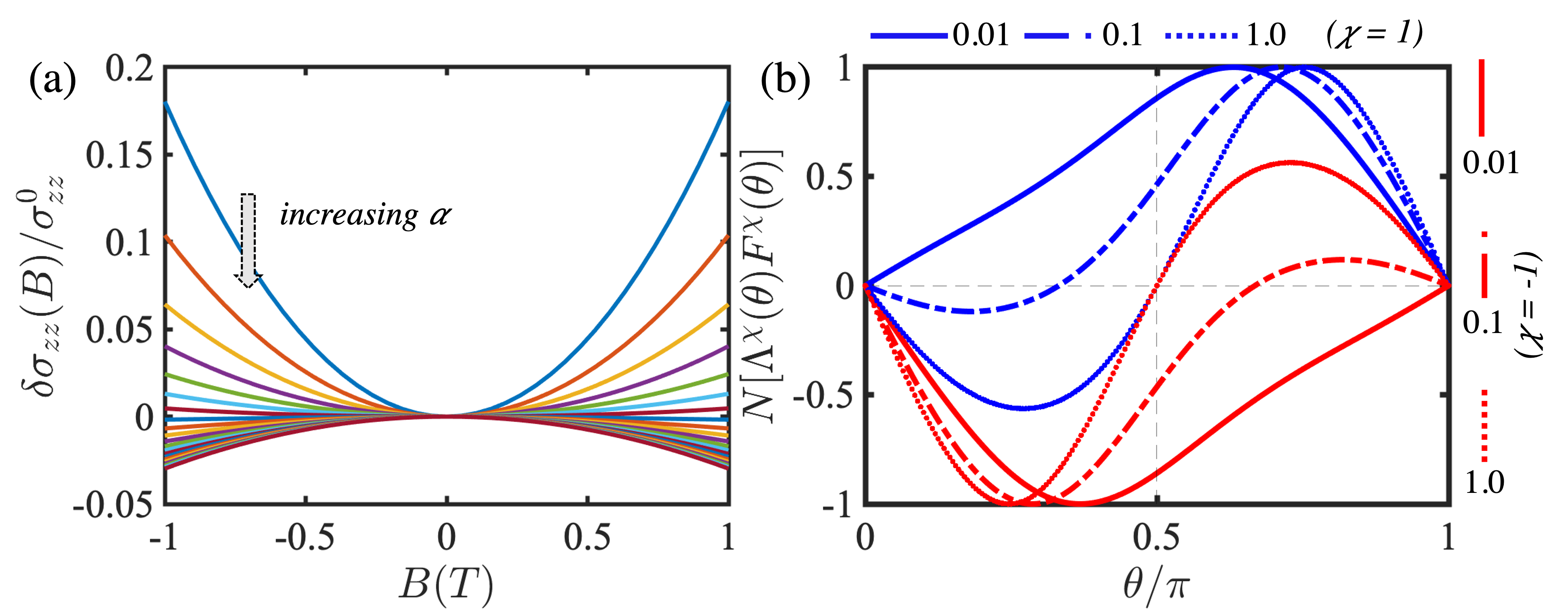}
    \caption{(a) LMC for a pair of Weyl nodes of opposite chiralities becomes negative beyond a critical intervalley scattering strength $\alpha_c$. Here $\alpha$ is increased from 0.1 to 1. (b) The normalized deviation in distribution functions (proportional to $\delta f^\chi_\mathbf{k}$) at both the valleys for three different values of $\alpha$. }
    \label{fig:two_node_1}
\end{figure}
\begin{figure}
    \centering
    \includegraphics[width=\columnwidth]{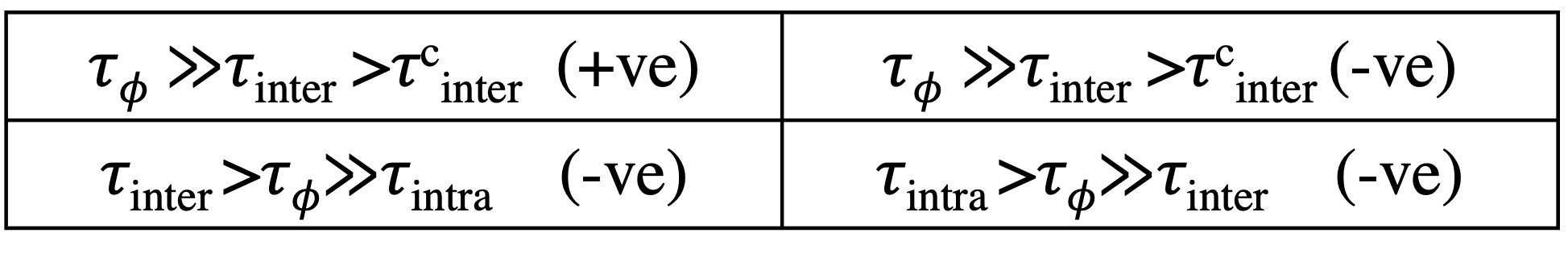}
    \caption{Conditions to observe positive or negative LMC for TR-broken Weyl semimetals. }
    \label{fig:table}
\end{figure}
\begin{figure}
    \centering
    \includegraphics[width=\columnwidth]{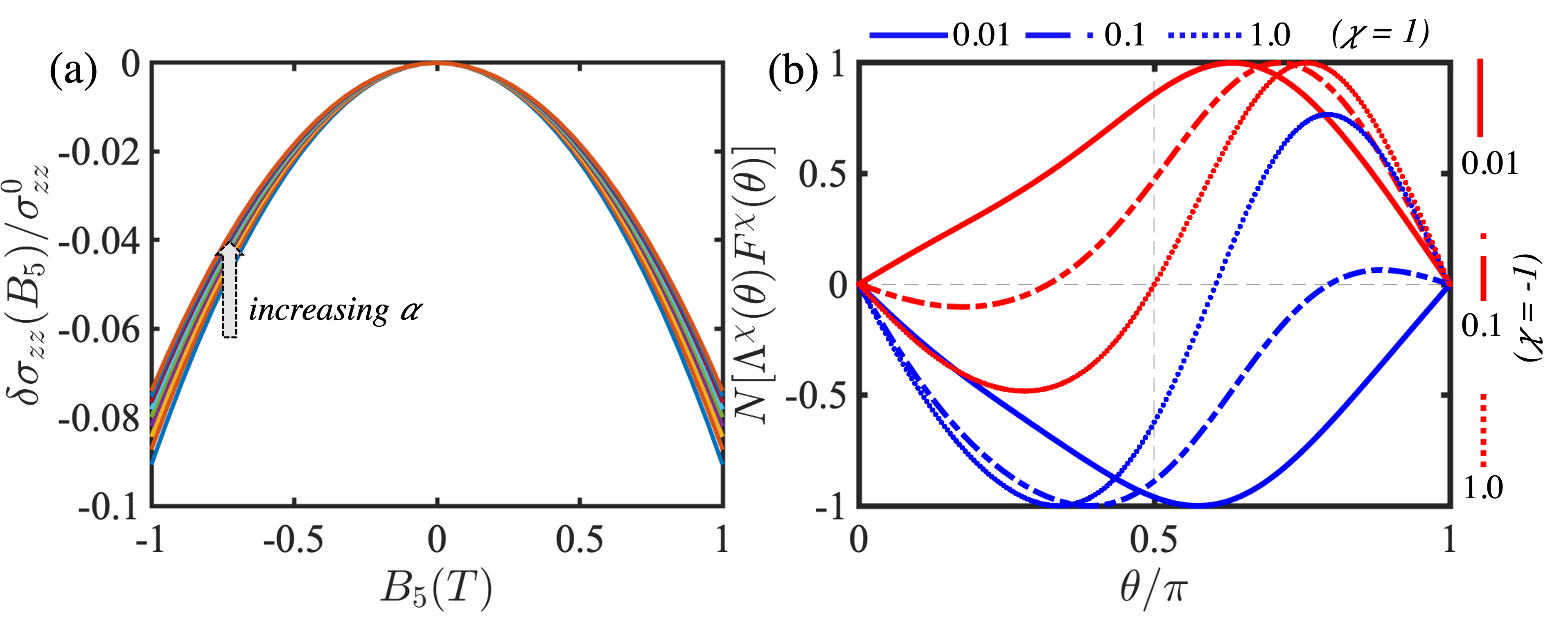}
    \caption{(a) Strain induced LMC for a pair of Weyl nodes of opposite chiralities is always negative. Here $\alpha$ is increased from 0.1 to 1. (b) The normalized deviation in distribution functions (proportional to $\delta f^\chi_\mathbf{k}$) at both the valleys for three different values of $\alpha$. $B_5$ field is chosen parallel to $\mathbf{E}$. }
    \label{fig:two_nodes_strain_1}
\end{figure}
Fig.~\ref{fig:one_node_1} (a) plots the evaluated LMC, $\delta\sigma_{zz}(B)$
for an isolated Weyl node, which turns out to be always negative, and independent of the chirality of the Weyl node. This result is in direct contrast to the recent papers concluding that intranode scattering alone can produce positive LMC in WSMs~\cite{kim2014boltzmann,lundgren2014thermoelectric,cortijo2016linear,sharma2016nernst,zyuzin2017magnetotransport,knoll2020negative,das2019berry,kundu2020magnetotransport}. Fig.~\ref{fig:one_node_1} (b) plots $\Lambda^\chi(\theta)$ as a function of the polar angle $\theta$, while Fig.~\ref{fig:one_node_1} (c) plots the normalized distribution function. The distribution function directly reflects $\delta f^\chi_\mathbf{k}$, which integrates to zero, as expected from particle number conservation. Since we assume zero intervalley scattering, our result is valid in the limit when two Weyl nodes that are practically uncoupled. In other words the internode scattering time $\tau_\mathrm{inter}$ is largest of all relevant timescales. As a complementary approach, we also solve this problem using an ansatz free, numerical solution to the Boltzmann equation, and reproduce the same findings~\cite{SI}. 

\textit{LMC with non-zero internode scattering:}
The collision integral takes the following form 
\begin{align}
    \mathcal{I}_\mathrm{coll}\{f_\mathbf{k}^\chi\}=\sum\limits_{\mathbf{k}'}\sum\limits_{\chi'} \left(\Lambda^{\chi}_\mathbf{k} - \Lambda^{\chi'}_{\mathbf{k}'}\right) W^{\chi\chi'}_{\mathbf{k}\mathbf{k}'},
\end{align}
where the scattering rate $W^{\chi\chi'}_{\mathbf{k}\mathbf{k}'}$ has internode as well as intranode scattering. Again, for point-like disorder the scattering rate is evaluated to be
\begin{align}
    W^{\chi\chi'}_{\mathbf{k}\mathbf{k}'} &= \frac{2\pi}{\hbar} |U^{\chi\chi'}|^2 \delta(\epsilon_\mathbf{k} - \epsilon_F) \times \nonumber\\&(1+\chi\chi'(\cos\theta\cos\theta' + \sin\theta\sin\theta' \cos(\phi-\phi'))),
    \label{Eq_W_2}
\end{align}
where the disorder parameter $U^{\chi\chi'}$ allows us to tune the internode and intranode scattering strengths differently. Hereafter, we denote the ratio $|U^{\chi\neq\chi'}|^2/ |U^{\chi=\chi'}|^2 \equiv \alpha$.
The solution presented before is extended for the case of two nodes~\cite{SI} with finite internode scattering. Fig.~\ref{fig:two_node_1} (a) plots LMC for increasing intervalley scattering strength. We find that LMC for a pair of Weyl nodes of opposite chiralities becomes negative beyond a critical intervalley scattering strength $\alpha_c$, as also observed in Ref.~\cite{knoll2020negative, sharma2020sign}. This is attributed to orbital magnetic moment that has opposite effect on the Fermi surfaces of both nodes~\cite{knoll2020negative}, and renders them dissimilar in the presence of external magnetic field.

The striking observation here is that even for weak internode scattering, LMC is positive, as opposed to the case of strictly zero internode scattering. We also note that the distribution function obtained here for the case of weak internode scattering is very different from the one obtained in Fig.~\ref{fig:one_node_1} for the case of zero internode scattering. In the latter case the deviation of the distribution function must integrate out to zero at a single node, i.e., the chiral charge is conserved. This condition is relaxed in the presence of weak internode scattering, as particles can scatter across nodes, and only the global charge  needs to be conserved. 

It must be noted that the Boltzmann transport equation gives the steady-state solution that is valid in the limit $t\gg \max\{\tau_{\mathrm{inter}}, \tau_{\mathrm{intra}}\}$. Hence, the particles are allowed to scatter and redistribute across both the nodes on a timescale much less than $\tau_\phi$. 
These observations summon  very important points: (i) intranode scattering, by itself, does not yield positive LMC due to chiral charge conservation, (ii) finite internode scattering is a \textit{necessary} condition for observing positive LMC even in the semiclassical low-$B$ limit, (iii) chiral anomaly induced positive LMC is therefore a pure internode phenomenon, reconciling the Boltzmann and the Landau-level picture, (iv) sufficiently large internode scattering beyond a critical value $\alpha_c$ switches the sign of LMC from positive to negative.
Correspondingly, $\tau_\phi\gg\tau_\mathrm{inter}>\tau_\mathrm{inter}^\mathrm{c}$ is necessary to observe positive LMC in experiments, where $\tau_\mathrm{inter}^\mathrm{c}$ is the critical intervalley scattering time below which LMC becomes negative. %Now, on increasing $\tau_\mathrm{intra}$ beyond a critical intervalley scattering time $\tau^\mathrm{c}_\mathrm{inter}$, LMC becomes negative. 
On the other hand, when $\tau_\mathrm{inter}$ becomes less than $\tau_\phi$, one must always observe negative LMC as dictated by $\tau_\mathrm{intra}$.
Fig.~\ref{fig:table} summarizes the conditions to observe positive and negative LMC.
\begin{figure}
    \centering
    \includegraphics[width=\columnwidth]{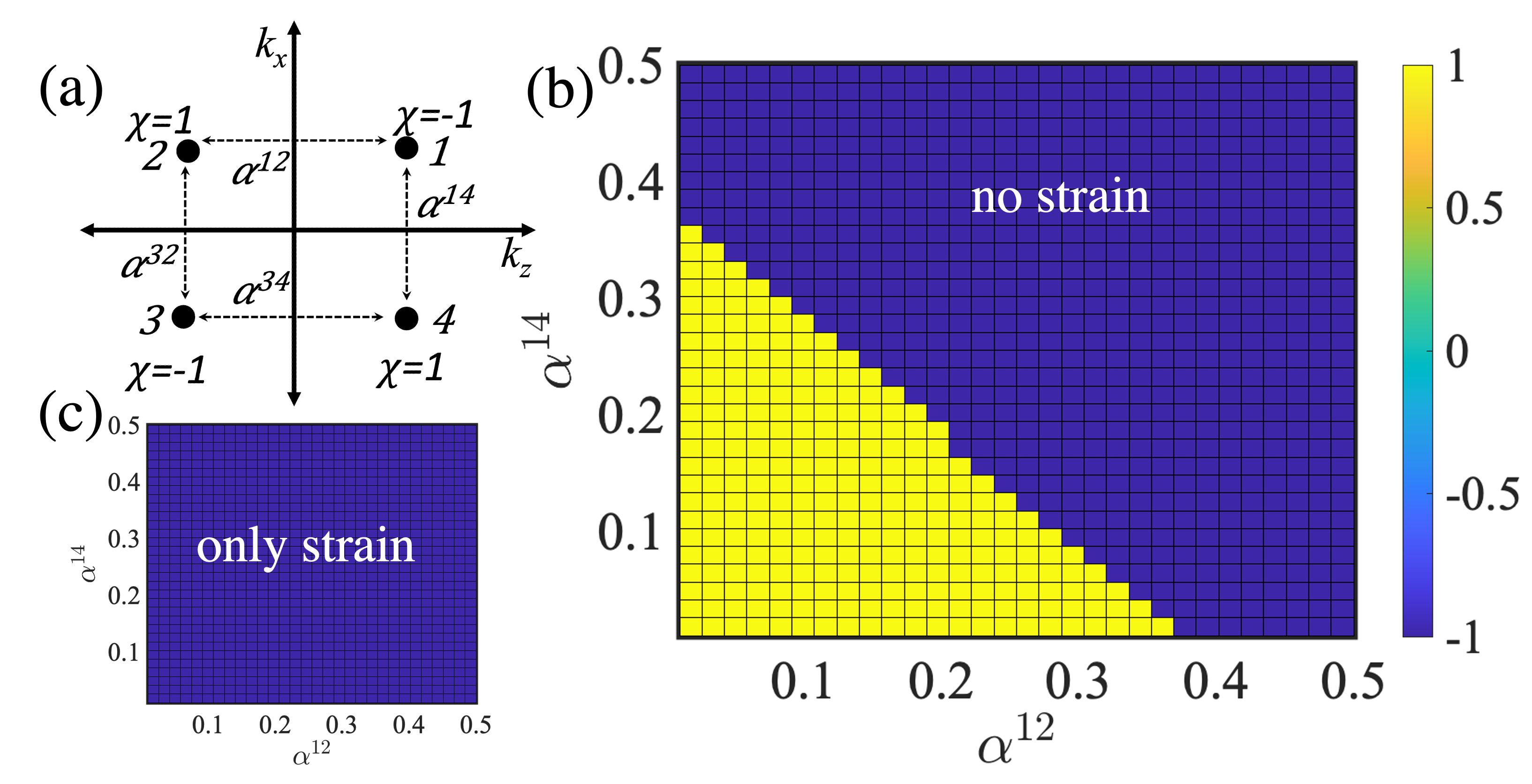}
    \caption{(a) Minimal model of inversion asymmetric WSM with four nodes. Chirality of a node is indicated by $\chi$, and internode scattering from node $i$ to $j$ (and vice-versa) is denoted by $\alpha^{ij}$. The sign of LMC as a function of internode scattering without strain is plotted in (b), and, with only strain in (c). Strain, alone, does not increase LMC. }
    \label{fig:four_nodes_1}
\end{figure}
\begin{figure}
    \centering
    \includegraphics[width=\columnwidth]{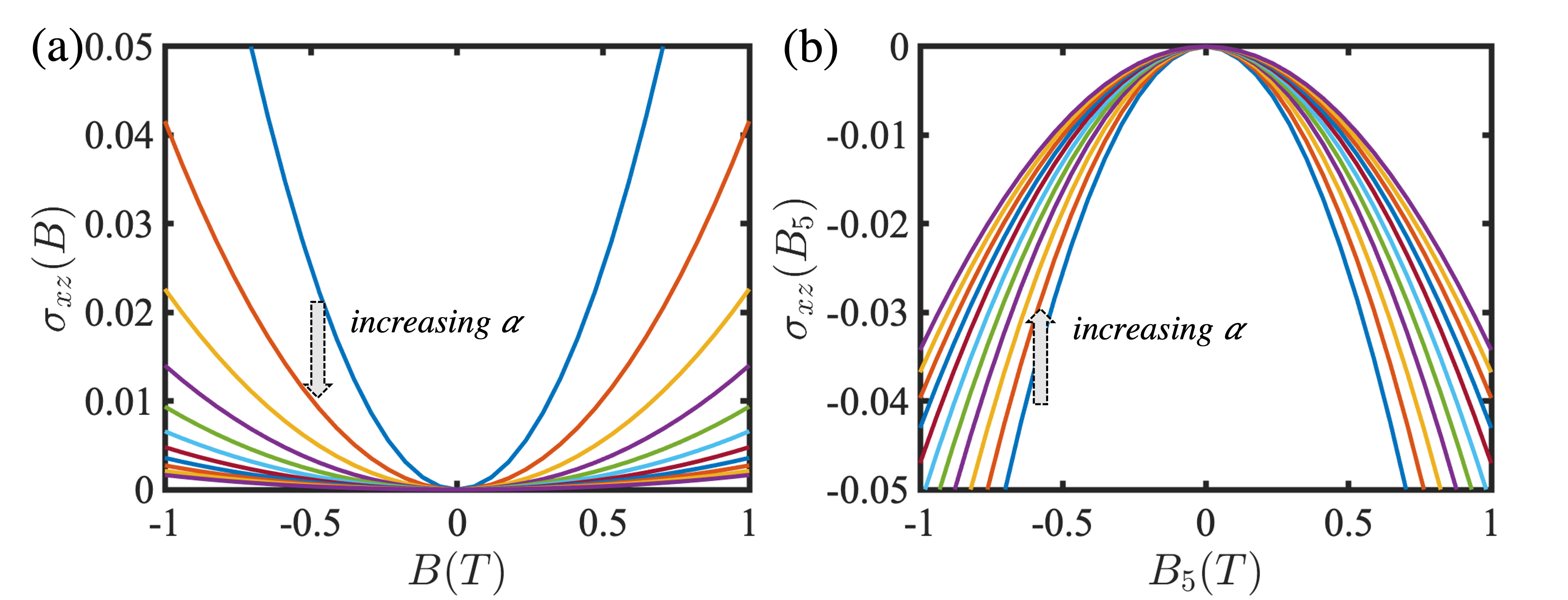}
    \caption{(a) Planar Hall conductivity as a function of magnetic field in the absence of strain, and, (b) as a function of the axial magnetic field $B_5$ in the absence of external magnetic field. The angle of the magnetic field in (a) and axial field (b) were chosen to be the same.}
    \label{fig:two_nodes_phe_1}
\end{figure}

\textit{Inhomogeneous WSMs:} We now show that our results have very important consequences in the characterization of Weyl systems with naturally occurring inhomogeneities. An axial magnetic field $\mathbf{B}_5$ that couples to Weyl fermions of opposite chirality with an opposite sign, can be realized by an inhomogeneous strain or magnetization profile in IWSMs~\cite{cortijo2015elastic,pikulin2016chiral,grushin2016inhomogeneous}. Thus, the magnetic field felt by each Weyl node is $\mathbf{B}^\chi = \mathbf{B}+\chi\mathbf{B}_5$. Interestingly, it has been pointed out that strain alone can result in positive contribution to LMC even in the absence of external magnetic field ($\mathbf{B}=0$)~\cite{grushin2016inhomogeneous}. In striking contrast, again, we discover that this result is incorrect, and strain alone does not yield positive LMC, but rather decreases conductance, as plotted in Fig.~\ref{fig:two_nodes_strain_1} (a). Furthermore, the distribution functions switch sign as compared to Fig.~\ref{fig:two_node_1}. This again is attributed to a combination of directional dependent scattering processes, as well as global charge conservation~\cite{SI}. In the absence of strain, the switching of LMC from positive to negative is attributed to the effect of orbital magnetic moment, but here, even ignoring the OMM contribution does not affect our result qualitatively. This is because, unlike the former case, strain induced OMM affects both the nodes equally, and thus the Fermi surfaces retain their similarity. 
We extend the formalism to the case of an inversion asymmetric WSM. We choose a prototype model of a system with four Weyl nodes located at the points $\mathbf{K}=(\pm \pi/2,0,\pm \pi/2)$ in the Brillouin zone.
The minimal model is given by 
\begin{align}
    H = \sum\limits_{n=1}^4 \chi_{n}\hbar v_F \mathbf{k}\cdot\boldsymbol{\sigma},
    \label{eq_H2nodes}
\end{align}
where $\chi_n$ is the chirality of each Weyl node, and specifically, we choose $\chi_1=-\chi_2=\chi_3=-\chi_4=-1$, such that time-reversal symmetry is respected (see Fig.~\ref{fig:four_nodes_1} (a)). We consider four possible internode scatterings as shown in the figure, and scattering between diagonal nodes is neglected. Furthermore, we choose $\alpha^{12}=\alpha^{34}$, and $\alpha^{32}=\alpha^{14}$.
The behavior of LMC without strain is plotted in Fig.~\ref{fig:four_nodes_1} (b). Sufficiently large internode scattering across (either $\alpha^{12}$ or $\alpha^{14}$) yields negative LMC. In the presence of only strain induced field $B_5$, we never get a positive contribution to LMC as shown in Fig.~\ref{fig:four_nodes_1} (c). 

\textit{Planar Hall effect in IWSMs:} To evaluate the planar Hall conductivity, the magnetic field is made non-collinear with the electric field direction. We rotate the magnetic field along the $x-z$ plane, that breaks the azimuthal symmetry. Nevertheless, the Boltzmann formalism can be generalized~\cite{SI} to evaluate the planar Hall conductivity. Focusing on the strain induced planar Hall effect, we find that contrary to previous reports~\cite{ghosh2020chirality}, the strain induced contribution is not only opposite to that of the regular planar Hall effect, but also is different in  magnitude. In Fig.~\ref{fig:two_nodes_phe_1} we explicitly show that for the same orientation of the $B$ and the $B_5$ field, the magnitude and the sign of planar Hall conductance are different from each other. Strain, thus has a   planar Hall conductance of opposite sign to that due to an external magnetic field. 

\textit{Final words:} Reconciling with the Landau-level picture, we show that chiral anomaly induced positive LMC is indeed an internode phenomenon, even in the weak-$B$ semiclassical limit. Our results have remarkable implications in the context of Weyl semimetals, including the result that strain induced axial magnetic field $B_5$, produced negative LMC, and contributes against the the planar Hall conductance due to an external magnetic field.

\textit{Acknowledgement:} G. S. acknowledges support from SERB Grant No.
IITM/SERB/GS/305. S. T. acknowledges support from Grant No. NSF 2014157.

\bibliography{biblio.bib}

\begin{thebibliography}{29}%
\makeatletter
\providecommand \@ifxundefined [1]{%
 \@ifx{#1\undefined}
}%
\providecommand \@ifnum [1]{%
 \ifnum #1\expandafter \@firstoftwo
 \else \expandafter \@secondoftwo
 \fi
}%
\providecommand \@ifx [1]{%
 \ifx #1\expandafter \@firstoftwo
 \else \expandafter \@secondoftwo
 \fi
}%
\providecommand \natexlab [1]{#1}%
\providecommand \enquote  [1]{``#1''}%
\providecommand \bibnamefont  [1]{#1}%
\providecommand \bibfnamefont [1]{#1}%
\providecommand \citenamefont [1]{#1}%
\providecommand \href@noop [0]{\@secondoftwo}%
\providecommand \href [0]{\begingroup \@sanitize@url \@href}%
\providecommand \@href[1]{\@@startlink{#1}\@@href}%
\providecommand \@@href[1]{\endgroup#1\@@endlink}%
\providecommand \@sanitize@url [0]{\catcode `\\12\catcode `\$12\catcode
  `\&12\catcode `\#12\catcode `\^12\catcode `\_12\catcode `\%12\relax}%
\providecommand \@@startlink[1]{}%
\providecommand \@@endlink[0]{}%
\providecommand \url  [0]{\begingroup\@sanitize@url \@url }%
\providecommand \@url [1]{\endgroup\@href {#1}{\urlprefix }}%
\providecommand \urlprefix  [0]{URL }%
\providecommand \Eprint [0]{\href }%
\providecommand \doibase [0]{https://doi.org/}%
\providecommand \selectlanguage [0]{\@gobble}%
\providecommand \bibinfo  [0]{\@secondoftwo}%
\providecommand \bibfield  [0]{\@secondoftwo}%
\providecommand \translation [1]{[#1]}%
\providecommand \BibitemOpen [0]{}%
\providecommand \bibitemStop [0]{}%
\providecommand \bibitemNoStop [0]{.\EOS\space}%
\providecommand \EOS [0]{\spacefactor3000\relax}%
\providecommand \BibitemShut  [1]{\csname bibitem#1\endcsname}%
\let\auto@bib@innerbib\@empty
%</preamble>
\bibitem [{\citenamefont {Adler}(1969)}]{adler1969axial}%
  \BibitemOpen
  \bibfield  {author} {\bibinfo {author} {\bibfnamefont {S.~L.}\ \bibnamefont
  {Adler}},\ }\bibfield  {title} {\bibinfo {title} {Axial-vector vertex in
  spinor electrodynamics},\ }\href@noop {} {\bibfield  {journal} {\bibinfo
  {journal} {Physical Review}\ }\textbf {\bibinfo {volume} {177}},\ \bibinfo
  {pages} {2426} (\bibinfo {year} {1969})}\BibitemShut {NoStop}%
\bibitem [{\citenamefont {Bell}\ and\ \citenamefont
  {Jackiw}(1969)}]{bell1969pcac}%
  \BibitemOpen
  \bibfield  {author} {\bibinfo {author} {\bibfnamefont {J.~S.}\ \bibnamefont
  {Bell}}\ and\ \bibinfo {author} {\bibfnamefont {R.}~\bibnamefont {Jackiw}},\
  }\bibfield  {title} {\bibinfo {title} {A pcac puzzle: $\pi$ 0→
  $\gamma$$\gamma$ in the $\sigma$-model},\ }\href@noop {} {\bibfield
  {journal} {\bibinfo  {journal} {Il Nuovo Cimento A (1965-1970)}\ }\textbf
  {\bibinfo {volume} {60}},\ \bibinfo {pages} {47} (\bibinfo {year}
  {1969})}\BibitemShut {NoStop}%
\bibitem [{\citenamefont {Armitage}\ \emph {et~al.}(2018)\citenamefont
  {Armitage}, \citenamefont {Mele},\ and\ \citenamefont
  {Vishwanath}}]{armitage2018weyl}%
  \BibitemOpen
  \bibfield  {author} {\bibinfo {author} {\bibfnamefont {N.}~\bibnamefont
  {Armitage}}, \bibinfo {author} {\bibfnamefont {E.}~\bibnamefont {Mele}},\
  and\ \bibinfo {author} {\bibfnamefont {A.}~\bibnamefont {Vishwanath}},\
  }\bibfield  {title} {\bibinfo {title} {Weyl and dirac semimetals in
  three-dimensional solids},\ }\href@noop {} {\bibfield  {journal} {\bibinfo
  {journal} {Reviews of Modern Physics}\ }\textbf {\bibinfo {volume} {90}},\
  \bibinfo {pages} {015001} (\bibinfo {year} {2018})}\BibitemShut {NoStop}%
\bibitem [{\citenamefont {Volovik}(2003)}]{volovik2003universe}%
  \BibitemOpen
  \bibfield  {author} {\bibinfo {author} {\bibfnamefont {G.~E.}\ \bibnamefont
  {Volovik}},\ }\href@noop {} {\emph {\bibinfo {title} {The universe in a
  helium droplet}}},\ Vol.\ \bibinfo {volume} {117}\ (\bibinfo  {publisher}
  {Oxford University Press on Demand},\ \bibinfo {year} {2003})\BibitemShut
  {NoStop}%
\bibitem [{\citenamefont {Nielsen}\ and\ \citenamefont
  {Ninomiya}(1981)}]{nielsen1981no}%
  \BibitemOpen
  \bibfield  {author} {\bibinfo {author} {\bibfnamefont {H.~B.}\ \bibnamefont
  {Nielsen}}\ and\ \bibinfo {author} {\bibfnamefont {M.}~\bibnamefont
  {Ninomiya}},\ }\href@noop {} {\emph {\bibinfo {title} {No-go theorum for
  regularizing chiral fermions}}},\ \bibinfo {type} {Tech. Rep.}\ (\bibinfo
  {institution} {Science Research Council},\ \bibinfo {year}
  {1981})\BibitemShut {NoStop}%
\bibitem [{\citenamefont {Nielsen}\ and\ \citenamefont
  {Ninomiya}(1983)}]{nielsen1983adler}%
  \BibitemOpen
  \bibfield  {author} {\bibinfo {author} {\bibfnamefont {H.~B.}\ \bibnamefont
  {Nielsen}}\ and\ \bibinfo {author} {\bibfnamefont {M.}~\bibnamefont
  {Ninomiya}},\ }\bibfield  {title} {\bibinfo {title} {The adler-bell-jackiw
  anomaly and weyl fermions in a crystal},\ }\href@noop {} {\bibfield
  {journal} {\bibinfo  {journal} {Physics Letters B}\ }\textbf {\bibinfo
  {volume} {130}},\ \bibinfo {pages} {389} (\bibinfo {year}
  {1983})}\BibitemShut {NoStop}%
\bibitem [{\citenamefont {Wan}\ \emph {et~al.}(2011)\citenamefont {Wan},
  \citenamefont {Turner}, \citenamefont {Vishwanath},\ and\ \citenamefont
  {Savrasov}}]{wan2011topological}%
  \BibitemOpen
  \bibfield  {author} {\bibinfo {author} {\bibfnamefont {X.}~\bibnamefont
  {Wan}}, \bibinfo {author} {\bibfnamefont {A.~M.}\ \bibnamefont {Turner}},
  \bibinfo {author} {\bibfnamefont {A.}~\bibnamefont {Vishwanath}},\ and\
  \bibinfo {author} {\bibfnamefont {S.~Y.}\ \bibnamefont {Savrasov}},\
  }\bibfield  {title} {\bibinfo {title} {Topological semimetal and fermi-arc
  surface states in the electronic structure of pyrochlore iridates},\
  }\href@noop {} {\bibfield  {journal} {\bibinfo  {journal} {Physical Review
  B}\ }\textbf {\bibinfo {volume} {83}},\ \bibinfo {pages} {205101} (\bibinfo
  {year} {2011})}\BibitemShut {NoStop}%
\bibitem [{\citenamefont {Xu}\ \emph {et~al.}(2011)\citenamefont {Xu},
  \citenamefont {Weng}, \citenamefont {Wang}, \citenamefont {Dai},\ and\
  \citenamefont {Fang}}]{xu2011chern}%
  \BibitemOpen
  \bibfield  {author} {\bibinfo {author} {\bibfnamefont {G.}~\bibnamefont
  {Xu}}, \bibinfo {author} {\bibfnamefont {H.}~\bibnamefont {Weng}}, \bibinfo
  {author} {\bibfnamefont {Z.}~\bibnamefont {Wang}}, \bibinfo {author}
  {\bibfnamefont {X.}~\bibnamefont {Dai}},\ and\ \bibinfo {author}
  {\bibfnamefont {Z.}~\bibnamefont {Fang}},\ }\bibfield  {title} {\bibinfo
  {title} {Chern semimetal and the quantized anomalous hall effect in hgcr 2 se
  4},\ }\href@noop {} {\bibfield  {journal} {\bibinfo  {journal} {Physical
  Review Letters}\ }\textbf {\bibinfo {volume} {107}},\ \bibinfo {pages}
  {186806} (\bibinfo {year} {2011})}\BibitemShut {NoStop}%
\bibitem [{\citenamefont {Zyuzin}\ \emph {et~al.}(2012)\citenamefont {Zyuzin},
  \citenamefont {Wu},\ and\ \citenamefont {Burkov}}]{zyuzin2012weyl}%
  \BibitemOpen
  \bibfield  {author} {\bibinfo {author} {\bibfnamefont {A.}~\bibnamefont
  {Zyuzin}}, \bibinfo {author} {\bibfnamefont {S.}~\bibnamefont {Wu}},\ and\
  \bibinfo {author} {\bibfnamefont {A.}~\bibnamefont {Burkov}},\ }\bibfield
  {title} {\bibinfo {title} {Weyl semimetal with broken time reversal and
  inversion symmetries},\ }\href@noop {} {\bibfield  {journal} {\bibinfo
  {journal} {Physical Review B}\ }\textbf {\bibinfo {volume} {85}},\ \bibinfo
  {pages} {165110} (\bibinfo {year} {2012})}\BibitemShut {NoStop}%
\bibitem [{\citenamefont {Son}\ and\ \citenamefont
  {Spivak}(2013)}]{son2013chiral}%
  \BibitemOpen
  \bibfield  {author} {\bibinfo {author} {\bibfnamefont {D.}~\bibnamefont
  {Son}}\ and\ \bibinfo {author} {\bibfnamefont {B.}~\bibnamefont {Spivak}},\
  }\bibfield  {title} {\bibinfo {title} {Chiral anomaly and classical negative
  magnetoresistance of weyl metals},\ }\href@noop {} {\bibfield  {journal}
  {\bibinfo  {journal} {Physical Review B}\ }\textbf {\bibinfo {volume} {88}},\
  \bibinfo {pages} {104412} (\bibinfo {year} {2013})}\BibitemShut {NoStop}%
\bibitem [{\citenamefont {Goswami}\ and\ \citenamefont
  {Tewari}(2013)}]{goswami2013axionic}%
  \BibitemOpen
  \bibfield  {author} {\bibinfo {author} {\bibfnamefont {P.}~\bibnamefont
  {Goswami}}\ and\ \bibinfo {author} {\bibfnamefont {S.}~\bibnamefont
  {Tewari}},\ }\bibfield  {title} {\bibinfo {title} {Axionic field theory of
  (3+ 1)-dimensional weyl semimetals},\ }\href@noop {} {\bibfield  {journal}
  {\bibinfo  {journal} {Physical Review B}\ }\textbf {\bibinfo {volume} {88}},\
  \bibinfo {pages} {245107} (\bibinfo {year} {2013})}\BibitemShut {NoStop}%
\bibitem [{\citenamefont {Goswami}\ \emph {et~al.}(2015)\citenamefont
  {Goswami}, \citenamefont {Pixley},\ and\ \citenamefont
  {Sarma}}]{goswami2015axial}%
  \BibitemOpen
  \bibfield  {author} {\bibinfo {author} {\bibfnamefont {P.}~\bibnamefont
  {Goswami}}, \bibinfo {author} {\bibfnamefont {J.}~\bibnamefont {Pixley}},\
  and\ \bibinfo {author} {\bibfnamefont {S.~D.}\ \bibnamefont {Sarma}},\
  }\bibfield  {title} {\bibinfo {title} {Axial anomaly and longitudinal
  magnetoresistance of a generic three-dimensional metal},\ }\href@noop {}
  {\bibfield  {journal} {\bibinfo  {journal} {Physical Review B}\ }\textbf
  {\bibinfo {volume} {92}},\ \bibinfo {pages} {075205} (\bibinfo {year}
  {2015})}\BibitemShut {NoStop}%
\bibitem [{\citenamefont {Zhong}\ \emph {et~al.}(2015)\citenamefont {Zhong},
  \citenamefont {Orenstein},\ and\ \citenamefont {Moore}}]{zhong2015optical}%
  \BibitemOpen
  \bibfield  {author} {\bibinfo {author} {\bibfnamefont {S.}~\bibnamefont
  {Zhong}}, \bibinfo {author} {\bibfnamefont {J.}~\bibnamefont {Orenstein}},\
  and\ \bibinfo {author} {\bibfnamefont {J.~E.}\ \bibnamefont {Moore}},\
  }\bibfield  {title} {\bibinfo {title} {Optical gyrotropy from axion
  electrodynamics in momentum space},\ }\href@noop {} {\bibfield  {journal}
  {\bibinfo  {journal} {Physical Review Letters}\ }\textbf {\bibinfo {volume}
  {115}},\ \bibinfo {pages} {117403} (\bibinfo {year} {2015})}\BibitemShut
  {NoStop}%
\bibitem [{\citenamefont {Kim}\ \emph {et~al.}(2014)\citenamefont {Kim},
  \citenamefont {Kim},\ and\ \citenamefont {Sasaki}}]{kim2014boltzmann}%
  \BibitemOpen
  \bibfield  {author} {\bibinfo {author} {\bibfnamefont {K.-S.}\ \bibnamefont
  {Kim}}, \bibinfo {author} {\bibfnamefont {H.-J.}\ \bibnamefont {Kim}},\ and\
  \bibinfo {author} {\bibfnamefont {M.}~\bibnamefont {Sasaki}},\ }\bibfield
  {title} {\bibinfo {title} {Boltzmann equation approach to anomalous transport
  in a weyl metal},\ }\href@noop {} {\bibfield  {journal} {\bibinfo  {journal}
  {Physical Review B}\ }\textbf {\bibinfo {volume} {89}},\ \bibinfo {pages}
  {195137} (\bibinfo {year} {2014})}\BibitemShut {NoStop}%
\bibitem [{\citenamefont {Lundgren}\ \emph {et~al.}(2014)\citenamefont
  {Lundgren}, \citenamefont {Laurell},\ and\ \citenamefont
  {Fiete}}]{lundgren2014thermoelectric}%
  \BibitemOpen
  \bibfield  {author} {\bibinfo {author} {\bibfnamefont {R.}~\bibnamefont
  {Lundgren}}, \bibinfo {author} {\bibfnamefont {P.}~\bibnamefont {Laurell}},\
  and\ \bibinfo {author} {\bibfnamefont {G.~A.}\ \bibnamefont {Fiete}},\
  }\bibfield  {title} {\bibinfo {title} {Thermoelectric properties of weyl and
  dirac semimetals},\ }\href@noop {} {\bibfield  {journal} {\bibinfo  {journal}
  {Physical Review B}\ }\textbf {\bibinfo {volume} {90}},\ \bibinfo {pages}
  {165115} (\bibinfo {year} {2014})}\BibitemShut {NoStop}%
\bibitem [{\citenamefont {Cortijo}(2016)}]{cortijo2016linear}%
  \BibitemOpen
  \bibfield  {author} {\bibinfo {author} {\bibfnamefont {A.}~\bibnamefont
  {Cortijo}},\ }\bibfield  {title} {\bibinfo {title} {Linear magnetochiral
  effect in weyl semimetals},\ }\href@noop {} {\bibfield  {journal} {\bibinfo
  {journal} {Physical Review B}\ }\textbf {\bibinfo {volume} {94}},\ \bibinfo
  {pages} {241105} (\bibinfo {year} {2016})}\BibitemShut {NoStop}%
\bibitem [{\citenamefont {Sharma}\ \emph {et~al.}(2016)\citenamefont {Sharma},
  \citenamefont {Goswami},\ and\ \citenamefont {Tewari}}]{sharma2016nernst}%
  \BibitemOpen
  \bibfield  {author} {\bibinfo {author} {\bibfnamefont {G.}~\bibnamefont
  {Sharma}}, \bibinfo {author} {\bibfnamefont {P.}~\bibnamefont {Goswami}},\
  and\ \bibinfo {author} {\bibfnamefont {S.}~\bibnamefont {Tewari}},\
  }\bibfield  {title} {\bibinfo {title} {Nernst and magnetothermal conductivity
  in a lattice model of weyl fermions},\ }\href@noop {} {\bibfield  {journal}
  {\bibinfo  {journal} {Physical Review B}\ }\textbf {\bibinfo {volume} {93}},\
  \bibinfo {pages} {035116} (\bibinfo {year} {2016})}\BibitemShut {NoStop}%
\bibitem [{\citenamefont {Zyuzin}(2017)}]{zyuzin2017magnetotransport}%
  \BibitemOpen
  \bibfield  {author} {\bibinfo {author} {\bibfnamefont {V.~A.}\ \bibnamefont
  {Zyuzin}},\ }\bibfield  {title} {\bibinfo {title} {Magnetotransport of weyl
  semimetals due to the chiral anomaly},\ }\href@noop {} {\bibfield  {journal}
  {\bibinfo  {journal} {Physical Review B}\ }\textbf {\bibinfo {volume} {95}},\
  \bibinfo {pages} {245128} (\bibinfo {year} {2017})}\BibitemShut {NoStop}%
\bibitem [{\citenamefont {Das}\ and\ \citenamefont
  {Agarwal}(2019)}]{das2019berry}%
  \BibitemOpen
  \bibfield  {author} {\bibinfo {author} {\bibfnamefont {K.}~\bibnamefont
  {Das}}\ and\ \bibinfo {author} {\bibfnamefont {A.}~\bibnamefont {Agarwal}},\
  }\bibfield  {title} {\bibinfo {title} {Berry curvature induced thermopower in
  type-i and type-ii weyl semimetals},\ }\href@noop {} {\bibfield  {journal}
  {\bibinfo  {journal} {Physical Review B}\ }\textbf {\bibinfo {volume}
  {100}},\ \bibinfo {pages} {085406} (\bibinfo {year} {2019})}\BibitemShut
  {NoStop}%
\bibitem [{\citenamefont {Kundu}\ \emph {et~al.}(2020)\citenamefont {Kundu},
  \citenamefont {Siu}, \citenamefont {Yang},\ and\ \citenamefont
  {Jalil}}]{kundu2020magnetotransport}%
  \BibitemOpen
  \bibfield  {author} {\bibinfo {author} {\bibfnamefont {A.}~\bibnamefont
  {Kundu}}, \bibinfo {author} {\bibfnamefont {Z.~B.}\ \bibnamefont {Siu}},
  \bibinfo {author} {\bibfnamefont {H.}~\bibnamefont {Yang}},\ and\ \bibinfo
  {author} {\bibfnamefont {M.~B.}\ \bibnamefont {Jalil}},\ }\bibfield  {title}
  {\bibinfo {title} {Magnetotransport of weyl semimetals with tilted dirac
  cones},\ }\href@noop {} {\bibfield  {journal} {\bibinfo  {journal} {New
  Journal of Physics}\ }\textbf {\bibinfo {volume} {22}},\ \bibinfo {pages}
  {083081} (\bibinfo {year} {2020})}\BibitemShut {NoStop}%
\bibitem [{\citenamefont {Knoll}\ \emph {et~al.}(2020)\citenamefont {Knoll},
  \citenamefont {Timm},\ and\ \citenamefont {Meng}}]{knoll2020negative}%
  \BibitemOpen
  \bibfield  {author} {\bibinfo {author} {\bibfnamefont {A.}~\bibnamefont
  {Knoll}}, \bibinfo {author} {\bibfnamefont {C.}~\bibnamefont {Timm}},\ and\
  \bibinfo {author} {\bibfnamefont {T.}~\bibnamefont {Meng}},\ }\bibfield
  {title} {\bibinfo {title} {Negative longitudinal magnetoconductance at weak
  fields in weyl semimetals},\ }\href@noop {} {\bibfield  {journal} {\bibinfo
  {journal} {Physical Review B}\ }\textbf {\bibinfo {volume} {101}},\ \bibinfo
  {pages} {201402} (\bibinfo {year} {2020})}\BibitemShut {NoStop}%
\bibitem [{\citenamefont {Sharma}\ \emph {et~al.}(2020)\citenamefont {Sharma},
  \citenamefont {Nandy},\ and\ \citenamefont {Tewari}}]{sharma2020sign}%
  \BibitemOpen
  \bibfield  {author} {\bibinfo {author} {\bibfnamefont {G.}~\bibnamefont
  {Sharma}}, \bibinfo {author} {\bibfnamefont {S.}~\bibnamefont {Nandy}},\ and\
  \bibinfo {author} {\bibfnamefont {S.}~\bibnamefont {Tewari}},\ }\bibfield
  {title} {\bibinfo {title} {Sign of longitudinal magnetoconductivity and the
  planar hall effect in weyl semimetals},\ }\href@noop {} {\bibfield  {journal}
  {\bibinfo  {journal} {Physical Review B}\ }\textbf {\bibinfo {volume}
  {102}},\ \bibinfo {pages} {205107} (\bibinfo {year} {2020})}\BibitemShut
  {NoStop}%
\bibitem [{\citenamefont {Nandy}\ \emph {et~al.}(2017)\citenamefont {Nandy},
  \citenamefont {Sharma}, \citenamefont {Taraphder},\ and\ \citenamefont
  {Tewari}}]{nandy2017chiral}%
  \BibitemOpen
  \bibfield  {author} {\bibinfo {author} {\bibfnamefont {S.}~\bibnamefont
  {Nandy}}, \bibinfo {author} {\bibfnamefont {G.}~\bibnamefont {Sharma}},
  \bibinfo {author} {\bibfnamefont {A.}~\bibnamefont {Taraphder}},\ and\
  \bibinfo {author} {\bibfnamefont {S.}~\bibnamefont {Tewari}},\ }\bibfield
  {title} {\bibinfo {title} {Chiral anomaly as the origin of the planar hall
  effect in weyl semimetals},\ }\href@noop {} {\bibfield  {journal} {\bibinfo
  {journal} {Physical Review Letters}\ }\textbf {\bibinfo {volume} {119}},\
  \bibinfo {pages} {176804} (\bibinfo {year} {2017})}\BibitemShut {NoStop}%
\bibitem [{\citenamefont {Mahan}(2008)}]{mahan20089}%
  \BibitemOpen
  \bibfield  {author} {\bibinfo {author} {\bibfnamefont {G.~D.}\ \bibnamefont
  {Mahan}},\ }\href@noop {} {\emph {\bibinfo {title} {9. Many-Particle
  Systems}}}\ (\bibinfo  {publisher} {Princeton University Press},\ \bibinfo
  {year} {2008})\BibitemShut {NoStop}%
\bibitem [{\citenamefont {Grushin}\ \emph {et~al.}(2016)\citenamefont
  {Grushin}, \citenamefont {Venderbos}, \citenamefont {Vishwanath},\ and\
  \citenamefont {Ilan}}]{grushin2016inhomogeneous}%
  \BibitemOpen
  \bibfield  {author} {\bibinfo {author} {\bibfnamefont {A.~G.}\ \bibnamefont
  {Grushin}}, \bibinfo {author} {\bibfnamefont {J.~W.}\ \bibnamefont
  {Venderbos}}, \bibinfo {author} {\bibfnamefont {A.}~\bibnamefont
  {Vishwanath}},\ and\ \bibinfo {author} {\bibfnamefont {R.}~\bibnamefont
  {Ilan}},\ }\bibfield  {title} {\bibinfo {title} {Inhomogeneous weyl and dirac
  semimetals: Transport in axial magnetic fields and fermi arc surface states
  from pseudo-landau levels},\ }\href@noop {} {\bibfield  {journal} {\bibinfo
  {journal} {Physical Review X}\ }\textbf {\bibinfo {volume} {6}},\ \bibinfo
  {pages} {041046} (\bibinfo {year} {2016})}\BibitemShut {NoStop}%
\bibitem [{\citenamefont {Ghosh}\ \emph {et~al.}(2020)\citenamefont {Ghosh},
  \citenamefont {Sinha}, \citenamefont {Nandy},\ and\ \citenamefont
  {Taraphder}}]{ghosh2020chirality}%
  \BibitemOpen
  \bibfield  {author} {\bibinfo {author} {\bibfnamefont {S.}~\bibnamefont
  {Ghosh}}, \bibinfo {author} {\bibfnamefont {D.}~\bibnamefont {Sinha}},
  \bibinfo {author} {\bibfnamefont {S.}~\bibnamefont {Nandy}},\ and\ \bibinfo
  {author} {\bibfnamefont {A.}~\bibnamefont {Taraphder}},\ }\bibfield  {title}
  {\bibinfo {title} {Chirality-dependent planar hall effect in inhomogeneous
  weyl semimetals},\ }\href@noop {} {\bibfield  {journal} {\bibinfo  {journal}
  {Physical Review B}\ }\textbf {\bibinfo {volume} {102}},\ \bibinfo {pages}
  {121105} (\bibinfo {year} {2020})}\BibitemShut {NoStop}%
\bibitem [{SI(2022)}]{SI}%
  \BibitemOpen
  \bibfield  {title} {\bibinfo {title} {See supplementary information},\
  }\href@noop {} {\  (\bibinfo {year} {2022})}\BibitemShut {NoStop}%
\bibitem [{\citenamefont {Cortijo}\ \emph {et~al.}(2015)\citenamefont
  {Cortijo}, \citenamefont {Ferreir{\'o}s}, \citenamefont {Landsteiner},\ and\
  \citenamefont {Vozmediano}}]{cortijo2015elastic}%
  \BibitemOpen
  \bibfield  {author} {\bibinfo {author} {\bibfnamefont {A.}~\bibnamefont
  {Cortijo}}, \bibinfo {author} {\bibfnamefont {Y.}~\bibnamefont
  {Ferreir{\'o}s}}, \bibinfo {author} {\bibfnamefont {K.}~\bibnamefont
  {Landsteiner}},\ and\ \bibinfo {author} {\bibfnamefont {M.~A.}\ \bibnamefont
  {Vozmediano}},\ }\bibfield  {title} {\bibinfo {title} {Elastic gauge fields
  in weyl semimetals},\ }\href@noop {} {\bibfield  {journal} {\bibinfo
  {journal} {Physical Review Letters}\ }\textbf {\bibinfo {volume} {115}},\
  \bibinfo {pages} {177202} (\bibinfo {year} {2015})}\BibitemShut {NoStop}%
\bibitem [{\citenamefont {Pikulin}\ \emph {et~al.}(2016)\citenamefont
  {Pikulin}, \citenamefont {Chen},\ and\ \citenamefont
  {Franz}}]{pikulin2016chiral}%
  \BibitemOpen
  \bibfield  {author} {\bibinfo {author} {\bibfnamefont {D.}~\bibnamefont
  {Pikulin}}, \bibinfo {author} {\bibfnamefont {A.}~\bibnamefont {Chen}},\ and\
  \bibinfo {author} {\bibfnamefont {M.}~\bibnamefont {Franz}},\ }\bibfield
  {title} {\bibinfo {title} {Chiral anomaly from strain-induced gauge fields in
  dirac and weyl semimetals},\ }\href@noop {} {\bibfield  {journal} {\bibinfo
  {journal} {Physical Review X}\ }\textbf {\bibinfo {volume} {6}},\ \bibinfo
  {pages} {041021} (\bibinfo {year} {2016})}\BibitemShut {NoStop}%
\end{thebibliography}%
\end{document}